\documentclass[12pt]{article}
\usepackage{epic,eepic}
\def\beq{\begin{equation}}
\def\eeq{\end{equation}}
\def\bea{\begin{eqnarray}}
\def\eea{\end{eqnarray}}

\def\ba{\begin{array}}
\def\ea{\end{array}}
\def\bce{\begin{center}}
\def\ece{\end{center}}

\begin{document}
\begin{titlepage}
\rightline{BROWN-HET-1162}
\rightline{hep-th/9812097}
\def\today{\ifcase\month\or
January\or February\or March\or April\or May\or June\or
July\or August\or September\or October\or November\or December\fi,
\number\year}
\vskip 1cm
\centerline{\Large \bf Moduli Space for Conifolds as Intersection of}
\centerline{\Large \bf Orthogonal D6 branes}
\vskip 1cm
\centerline{\sc  Robert de Mello Koch$^{b}$, Kyungho Oh$^{a}$ and Radu Tatar$^{b}$}
\vskip 1cm
\centerline{ $^a$ Dept. of Mathematics, University of 
Missouri-St. Louis,
St. Louis, MO 63121, USA }
\centerline{{\tt oh@math.umsl.edu}}
\centerline{$^b$ Dept. of Physics, Brown University,
Providence, RI 02912, USA}
\centerline{\tt robert,tatar@het.brown.edu}
\vskip 2cm
\centerline{\sc Abstract}
We  show that a system of parallel D3 branes near a conifold singularity can 
be mapped onto an intersecting configuration of orthogonal branes in type IIA 
string theory. Using this brane configuration, we analyze the Higgs moduli space
of the associated field theory. The dimension of the Higgs moduli space is
computed from a geometrical analysis of the conifold singularity. Our results
provide evidence for an extended s-rule. In addition, a discrepancy between
the prediction of the brane configuration and the result obtained from a 
geometrical analysis is noted. This discrepancy can be traced back to 
worldsheet instanton effects.
\vskip 0.2in

\end{titlepage}
\newpage
\section{Introduction}
\setcounter{equation}{0}
In the last few years it has become clear that gauge theory and gravity are 
complementary descriptions of a single theory. Insights which have
clarified and motivated these important ideas have largely been
developed using the solitonic brane solutions of M theory and string theory.
In particular, configurations containing NS5 fivebranes 
and D branes in string theory provide a useful way of studying supersymmetric 
gauge field theory in different dimensions and with different amounts of
unbroken supersymmetries (see \cite{gk} for a detailed review and with a 
complete set of references up to February 1998). As an example of the power of
the brane approach, we mention that these brane configurations 
provide techniques which may be used to derive large classes of Seiberg 
dualities (for ${\cal N}=1$ theories) and to obtain exact results after 
lifting to M theory (for both ${\cal N}=1,2$ theories)\cite{wit1}. Interesting
recent work in this field has focused on the study of D branes which are finite
in one direction \cite{ura2,kls,aot1,asu,park,ho1,lll,ls,cp,lo,kar}.
Another variant of
this approach is provided by Brane Boxes which allow the study of D branes which
are finite in two spacetime direction \cite{hz,gg,ab,lr,hu,rsu,klm}. These
studies
have provided new insights into finite gauge theories and chiral four dimensional
gauge theories among other things.

The use of brane configurations to describe the field theory realized on the
worldvolume of D branes on orbifold singularities has been given in
\cite{dgm,dm} (see \cite{hu} for a connection with Brane Box Models). The
study of these brane configurations is interesting because they provide
simple examples of field theories with a reduced amount of supersymmetry.
Recently, in an extension of these ideas, D branes on non-orbifold singularities
have been considered. The conifold singularity has been analyzed in \cite{kw} 
where an infrared theory on the worldvolume of D3 branes was proposed. This
provides a novel extension of the original AdS-CFT correspondence to a case 
with reduced supersymmetry and thus where the compact space is not even locally 
$S^{5}$. Other results for the case of non-orbifold singularities and their
connection to field theories in three and four dimensions have been obtained in 
\cite{afhs,gr1,mr,str,ot,bg,lopez}. 

Motivated by these preliminary studies, a more systematic way of studying
 D branes
in the presence of conifold singularities has been developed in \cite{ura,dm}.
These authors have exploited the fact that 
the conifold singularity is dual to a 
system of perpendicular NS5 fivebranes intersecting over a 3+1 dimensional
world-volume. 

In this paper we use a different approach to explore a system of D3 branes in
the geometry given by a general conifold $x y = v^m w^n$. In a recent 
paper \cite{dm}
the supergravity solution for intersecting branes \cite{bre} was used to give a
heuristic but rather explicit map from brane configurations to conifolds.
Motivated by this observation, we obtain new brane configurations together with
the corresponding conifolds. In particular, we
are able to describe a system of D3 branes at a conifold singularity in terms
of D4 branes in the presence of $m$ D6 branes (spanning the 0123789
directions) and $n$ D6$'$ branes (spanning the 0123457 directions). The 
study of
this brane configuration is interesting not only because it provides new and
more general maps between brane configurations and conifolds, but also because 
it involves fascinating worldsheet
instanton effects. These effects are non-perturbative in $l_{s}$ and are
still important in the limit $g_{s}\to 0$. They give rise to a long ranged
repulsive interaction between D4 branes stretched between 
a D6 and D6$'$ brane~\cite{gk}.
When constructing brane configurations, one imposes this rule in much the same
way that the s rule is imposed. One of the issues which we address is a 
possible geometric explanation for this rule, which is motivated by the natural
emergence of the s rule from purely geometrical considerations.
A generalization of the result obtained in \cite{wit1} for a single type of D6
brane shows that the configuration involving both D6 and D6$'$ branes gives 
rise
to the geometry with a singularity of the form $x y = v^m w^n$.
We will argue below that our approach yields the 
correct dimension of the Higgs moduli space. In this way,
we are able to generalize the result of \cite{hoo} where the dimension
of the Higgs moduli space was calculated by blowing-up the $A_{n-1}$
singularity and then counting the multiplicities of the ${\bf P^1}$
curves appearing in the resolved surface. In our case the details are more
involved because we need to resolve the $A_{m-1, n-1}$ singularity which
is a singularity of the form $x y = v^m w^n$. 

Our results in the case of configurations involving only D6 or D6$'$ branes
show complete agreement between the geometrical results and those predicted
by the brane configurations. Our geometrical analysis provides evidence that 
the s-rule continues to hold when the D6 and NS5 branes are are at any
non-zero angle with respect to each other. To the best of our knowledge,
this is a new result. In the case of configurations involving D6 and  D6$'$ 
branes we find an interesting discrepancy between the geometrical results as
compared to those predicted by the brane configurations.
This  is rather unexpected in light of previous  results
involving only D6 branes~\cite{wit1, hoo}. 
The discrepancy
can be traced back to the repulsive interaction between D4 branes stretched
between D6 and D6$'$ branes. Euclidean fundamental strings (i.e. worldsheet
instantons) are responsible for this repulsive interactions. 
Our  geometric model  is apparently not corrected by these 
instanton effects. Maybe the geometry must be modified
to reflect the  instanton effects.
We leave 
appropriate  treatments of these effects in the M theory framework as an 
interesting open question.

The remainder of the paper is organized as follows: In section 2 we begin
with an explanation of the set-up of our problem. Our starting point is
the observation that there is a direct connection between the metric
of a 3-brane at a conifold (obtained in equation (5. 13) of \cite{dm})
and the metric obtained from the supergravity solution for intersecting branes
(described in equation (20) of \cite{bre}),
obtained by performing a dimensional reduction. We then show that by
using (20) of \cite{bre} we can obtain a configuration with
two orthogonal D6 branes as opposed to the configuration with two orthogonal 
KK5 monopoles (which is of course connected
by a T-duality to a configuration with two orthogonal NS5 branes). 
In \cite{dm} the metric for two orthogonal NS5 branes was shown
to be the metric near the conifold singularity. This suggests that the metric for
two orthogonal D6 branes is also the metric near the conifold singularity.
This is not surprising and could have been anticipated from the results of \cite{wit1} 
where it was demonstrated that the presence of $m$ D6 branes leads naturally to the  
geometry $ x y = v^m$. The Higgs moduli space is discussed using a brane
configuration in type IIA theory in section 3. The brane configuration
consists of D4 branes suspended between NS5 branes, in the presence
of D6 and D6$'$ branes. The dimension of the
Higgs moduli space is computed by allowing the D4 branes to break into
segments suspended between NS5, D6 and D6$'$ branes. 
There are two distinct ``s-rules'' that
must be enforced in order to obtain the correct result. The first s-rule  
constrains the number of D4 branes stretched between NS5 and D6 branes.
The second s-rule is directly related to the worldsheet instanton effects
that we mentioned above. It constrains the number of D4 branes that can be
stretched between D6 and D6$'$ branes, in order to have
a stable brane configuration. Section 4 contains a discussion of relevant field theory 
results and their implications. In this section, we also show how to solve 
the simplest quadratic threefold singularity $x y = v w$.
In section 5 we solve the $A_{-1, n - 1}$ singularity which corresponds
to a configuration with D6$'$ branes and no D6 branes. In field theory this 
corresponds to an 
${\cal N} = 1$ theory with fundamental flavors and no superpotential.
Our results on the solution of the most general $A_{m-1, n-1}$ singularity
are presented in section 6. 

\section{Branes at threefold singularities}
\setcounter{equation}{0}
A useful approach to the study of conifolds has been developed in \cite{ura,dm}
where a conifold was mapped into a set of intersecting NS5 and NS5$'$ branes.  
The conifold is described in terms of two degenerating tori which vary
over a ${\bf P^1}$ base. The pair of orthogonal NS5 branes is obtained by
performing two T dualities. The first and second T dualities are performed 
along a cycle of the first and second tori. As discussed in \cite{ura,dm}, 
when the NS5 branes have three of their five dimensions common, they give rise 
to conifold singularities. The argument of \cite{dm} is directly relevant to
our study, and it is worth recalling some key points. These authors consider
a configuration involving NS5 branes (spanning the 012389 directions), NS5$'$
branes (spanning the 012345 directions) and D3 branes (spanning the 0126
directions). The supergravity metric for this brane configuration resembles 
the metric for a 3-brane at a conifold singularity\cite{dm}. Our arguments
will make use of the metric describing two Kaluza-Klein monopoles with a
five dimensional common worldvolume in 11 dimensions 
(see (20) of \cite{bre}). From the results of\cite{dm}, after removing the contribution of the D3 brane 
to the metric we have
\begin{equation}
\label{1}
ds^2 = ds_{0123}^2 + H_{5}' ds_{45}^2 + H_5 ds_{89}^{2} 
+ H_5 H_{5}' ds_7^2 + (H_5 H_{5}')^{-1} (ds_6 + A_1 d s_4 + B_1 ds_8)^2.
\end{equation}
Compare this to equation (20) of \cite{bre}
\begin{equation}
\label{2}
ds^2 = ds_{0123}^2 + ds_{10}^2 + H_{1} ds_{45}^2 + H_1 ds_{89}^{2}
+ H_1 H_2 ds_7^2 + (H_1 H_2)^{-1} (ds_6 + A_1 d s_4 + B_1 ds_8)^2
\end{equation}
where we have changed the notation in order to compare with the result of
\cite{dm}. A total of 6 gauge fields are considered in \cite{bre}; 
however, four of these can be
gauged to zero. All the harmonic functions in (\ref{1}) and
(\ref{2}) depend only on the $x^7$ coordinate. A key observation is that 
(\ref{1}) can be obtained from (\ref{2}) by reduction along $x^{10}$.
If we reduce along the $x^6$ direction, the resulting metric describes 
the intersection between a D6 brane (spanning the 123789 directions) and 
a D6$'$ brane (spanning the 123457 directions).  So the configuration with 
 D6(123789) and D6$'$(123457) and the one obtained from NS5-NS5$'$ after a
T-duality are both obtained from the same solution in 11 dimensions; the
only difference is that the dimensional reduction is performed on different 
directions in the two cases. This relationship between the two 
configurations, which has been argued at the level of supergravity, is
related to a duality between the two configurations at the level
of the full string theory. Both brane configurations correspond to a
single brane configuration from the point of view of M theory. The M theory
set up includes Kaluza Klein monopoles.
To obtain the IIA brane configuration which contains D6 branes, we
obtain IIA string theory from M theory by taking the strong coupling
eleventh dimension to be transverse to the 
M theory Kaluza Klein  monopoles. To obtain the IIA
configuration which contains Kaluza Klein monopoles, we take the strong
coupling direction to be parallel to the M theory Kaluza Klein monopoles.
Thus from the point of view of M theory, the two configurations are related
by a flip of the strong coupling direction\cite{jr,dvv} 
with another direction so that
the equivalence of these two configurations follows from eleven dimensional
Lorentz invariance.

This shows that system of D6 and D6$'$ branes 
gives a geometry which is similar to the geometry obtained from
the NS5-NS5$'$ configuration. An important result of \cite{dm} 
is that by T-dualizing a configuration with NS5 - NS5$'$ branes one maps
the metric to the metric of a conifold singularity:
\begin{equation}
\label{ref1}
x y = v w
\end{equation}
Our previous observation, lead us to the conclusion that the metric for a
configuration with D6 and D6$'$ branes can be mapped to (\ref{ref1}). 
With hindsight, we see that this geometry could have been anticipated 
from the results of \cite{wit1}. There it was shown that the presence of a 
D6 brane induces a change in the geometry of the form $x y = v$, whilst the 
presence of a D6$'$ brane would a change in the geometry of the form $x y = w$. 
In view of these results, it is natural to expect that one obtains $x y = v w$ 
in the presence of a system of D6 - D6$'$ branes. Our discussion above shows that 
this is indeed the geometry which appears. In \cite{wit1} the geometric 
structure corresponding to a system of n D6 branes on top of each other,
by using results of \cite{gr}. The geometric structure is 
that of a Taub - NUT space, which is a hyperK\"ahler manifold with three
complex structures. One of the complex structures is $x y = v^n$ i.e.
the $A_{n - 1}$ singularity. In our case, we do not have any localized
solution for the system of D6 - D6$'$ branes so we have been unable 
to obtain the complex structure directly from a specific metric. This is
why we have used the previous correspondence with a configuration of 
NS5 and NS5$'$ branes to identify the metric. 

The supersymmetry allows us to introduce in the final configuration two
types of NS5 branes, one in directions 012345 denoted by NS5 and one in the
012389 directions denoted by NS5$'$. In this way obtain a standard configuration
for ${\it N} = 1$ supersymmetric field theories in 4 dimensions 
\cite{egk,gk}. As usual strings between D4 branes give the gauge group
which is $SU(N)$ for a stack of N D4 branes, strings between
the D6 brane and D4 branes give one quark, denoted by $A$ and the 
strings
between the D6$'$ brane and the D4 branes give a second quark denoted
by $B$. 

The general case can be considered by using 
the following brane configuration:
Take $N_f$ parallel D6-branes extended
in the $(x_0, x_1, x_2, x_3, x_7, x_8, x_9)$ directions, located at
$d_i$ in $v$-plane and $N_f'$ parallel D6$'$-branes extended
in the $(x_0, x_1, x_2, x_3, x_4, x_5, x_7)$ directions, located at
$e_i$ in $w$-plane. In what follows, we
consider M theory on ${\bf R}^{10} \times {\bf S}^1$. This is equivalent
to Type IIA on ${\bf R}^{10}$, with the $U(1)$ gauge symmetry of
Type IIA being associated in M theory with the rotations of the
${\bf S}^1$. Via T and S dualities, as argued above, 
the D4 branes will live on a threefold given by 
\bea
xy = \prod_{i=1}^{N_f} (v -d_i) \prod_{i=1}^{N_f'} (w -e_i),
\eea
replacing the flat  $(x_4, x_5, x_6, x_7, x_8, x_{10})$ space due
the presence of D6 and D6$'$ branes.
Now, introduce two NS5 branes in
the $(x_0, x_1, x_2, x_3, x_4, x_5)$ directions and 
suspend $N_c$ D4 branes between them.
The equation of the Seiberg-Witten curve is given by 
\bea
& &w= 0 \\
& &x + y = B_{N_c}(v, u_k).
\eea

A more general form for the conifold can again be read from the results of
\cite{ura,dm} as:
\begin{equation}
x y = v^m w^n
\label{geom2}
\end{equation}
which can again be viewed as an orbifold of a ${\bf C^*}$ fibration over the 
${\bf C^2}$ parameterized by $v , w$ by ${\bf Z}_m \times {\bf Z}_n$. The 
T-dual configuration now contains $m$ NS5 branes spanning the
023457 directions, $n$ NS5 branes spanning the 023789 directions. After a
further T and S duality, we obtain $m$ D6 branes filling the 0123789 direction 
and $n$ D6$'$ branes filling the 0123457 directions. Finally, after introducing the 
NS5 and NS5$'$ brane, we obtain $m$ quarks $A_i$ from strings stretching between 
the D4 and D6 branes and $n$ quarks $B_j$
from strings stretching between the D4 and D6$'$ branes.
This is to be compared with the usual case when one has only one type of
D6 branes, giving only quarks of type A.

\section{Higgs Moduli space from Brane Configurations}
\setcounter{equation}{0}
In this section we review the description of the Higgs branch
in the type IIA picture. To reach the Higgs branch we allow the D4 branes
to break on the D6 and D6$'$ branes. After the breaking, 
there will be D4 branes
suspended between the D6 and D6$'$ branes. 
There are three distinct types of suspended D4 branes possible: a D4
brane suspended between a pair of D6 branes, a D4 brane suspended between 
a pair of D6$'$ branes and a D4 branes suspended between 
a D6 and a D6$'$ brane.
In addition to these, there are D4 branes between D6 and NS5 branes and 
D4 branes between D6$'$ branes and NS5 branes. The collective coordinates of a
D4-brane suspended between a pair of D6-branes consists of two complex
parameters built from the $x^7, x^8, x^9$ coordinate of the D4 together with 
the gauge field component $A_{6}$ corresponding to the compact $x^6$ coordinate.
Similarly, the location of a D4-brane between a pair D6$'$-branes is 
parameterized by two complex parameters built from the $x^4, x^5, x^7$ 
coordinate together with the gauge field component $A_{6}$. The location of 
a D4-brane between a D6 brane and a D6$'$ brane is parameterized by a single
complex coordinate built from the $x^7$ coordinate together with the gauge 
field component $A_{6}$. Finally, the location of a D4-brane between a D6$'$ 
and an NS5 brane is parameterized by one complex parameters built from the
$x^8, x^9$ coordinate of the brane. 
A D4 brane that is suspended between an NS5 brane and a D6 brane is not free
to move. 

When computing the dimension of moduli space, it is crucial that one
impose the s-rule which restricts the number of D4 branes stretched between
an NS5 brane and a D6 brane to one. For the configuration that we are studying 
there is an additional non-perturbative effect due to worldsheet instantons,
giving rise to a long range force between two D4 branes suspended between a D6
brane and a D6$'$ brane. This long-range force drives the D4 branes to infinity 
in the $x^7$ direction~\cite{gk}. Note that this effect is non-perturbative in $l_{s}$
and continues to remain important for the dynamics at arbitrarily weak string
coupling.

We can now proceed to calculate the dimension of the Higgs branch.In order to
have be able to compare with our results of section 6 we are going to 
discuss two cases, one with parallel NS branes in (12345) directions and one
with rotated NS branes in the (4589) plane. 
\subsection{Parallel NS Branes in the (12345) Directions}
 Consider
the case when we have $N_f$ D6$'$ branes. The theory has ${\cal N}=1$ supersymmetry.
The matter content of the theory includes $N_f$ fundamental flavors B (and 
their corresponding anti-fundamentals $\tilde{B}$) as well as an adjoint field
$\Phi$. In general, one expects a superpotential of the form
$\mbox{sin}(\theta) \tilde{B} \Phi B$ where $\theta$ is the angle between 
the D6 branes and the NS5 branes. In the case that we are considering, the D6$'$
branes are parallel to the NS5 branes so that the superpotential vanishes. 
The dimension of the Higgs moduli space is $2 N_f N_c$. Consider now the case 
of $N_f$ D6 branes. The brane configuration realizes a theory with 
${\cal N} = 2$ supersymmetry with a matter content the includes $N_f$ hypermultiplets in the
fundamental representation. It is well known that the dimension of the Higgs
moduli space is $2 N_f N_c - 2 N_c^2$. 

Let us turn now to a general configuration consisting of $n $
unrotated D6$'$ branes and $m$ rotated D6$'$ branes (i.e. D6 branes). There are
$N_c$ D4 branes suspended between the leftmost NS5 brane and the first
D6$'$ brane. These D4 branes make a contribution of $N_c$ to the
total (complex) dimension of the Higgs moduli space. There are $N_c$ D4 branes 
suspended between the D6$'$ branes which give a contribution of $2 (n - 1) N_c$ 
to the total complex dimension. Counting the number of D4 branes between the
D6 branes is a little more subtle because one has to correctly enforce the  s-rule. 
The s-rule places restrictions on the D4 branes connecting the
right most NS5 brane and the D6 branes. In addition, the worldsheet instanton
effects imply that a stable brane configuration is only obtained after
restricting to a single D4 brane between the rightmost D6$'$ brane and each of 
the D6 branes. Summing these contributions leads to a complex dimension of  
\begin{equation}
2 (m - 1) + 2 (m - 3) + \cdots 2 (m - 2 N_c + 1)=2 N_c (m - N_c).
\end{equation}
 The last thing contribution to the dimension of the Higgs moduli space
comes from the D4 branes suspended between the rightmost D6$'$ and the D6 
branes. This contribution is $N_c$. The final result is that the complex 
dimension of 
the Higgs moduli space is
\begin{equation}
\label{dim1}
2 (n - 1) N_c + 2 N_c ( m - N_c) + 2 N_c = 2 (n + m) N_c - 2N_c^2 = 
2 N_f N_c - 2 N_c^2.
\end{equation}
A comment is in order. Equation (\ref{dim1}) shows that the dimension of the
Higgs moduli space is the same as the dimension 
computed in the ${\cal N} = 2$ theory.
This is explained by noting that we can break the gauge symmetry by first 
introducing the D6 branes and only then introducing the D6$'$ branes to break 
the ${\cal N} = 2$ supersymmetry to 
${\cal N} = 1$. 
So, after introducing the D6 brane, the gauge symmetry is completely broken 
in the ${\cal N} = 2$ supersymmetric theory. The field theory calculations
tell us that the dimension of the Higgs moduli space is 
$2 m N_c - 2 N_c^2$. By introducing further flavors in the form of D6' branes
they will not break any more the gauge group and their contribution to the
Higgs moduli space is $2 m N_c$. Therefore, the field theory calculations
give the result $2 m N_c - 2 N_c^2 + 2 m N_c$ i.e. just the one of equation
(\ref{dim1}). 

The quaternionic dimension of the Higgs moduli space is 
\begin{equation}
\label{dim2}
 N_c (n + m) - N_c^2 .
\end{equation}
\subsection{Parallel NS Branes Rotated in the (4589) Plane}
Let us now discuss the case of $n$ unrotated D6$'$ branes and $m$ rotated D6$'$
branes (i.e. D6 branes) together with $N_c$ D4 branes suspended between
two parallel NS5 branes which are now at an angle in the (4589) plane.

In this case, the D4 branes stretched between NS5 
branes and D6 and D6$'$ branes are constrained by
the s-rule which restricts the number of D4 branes suspended 
between an NS5 brane and a D6 brane and an NS5 brane and a D6$'$ brane to one. 
There are several pieces of evidence for this extended s-rule. The neat fit
of the brane description with the field theory is only possible 
when this s-rule is enforced. 
The D4 branes suspended between a D6 brane
and a rotated NS5 brane or between a D6$'$ and a
rotated NS5 cannot move between them because they are extended in different 
directions. So these D4 branes would necessarily be on top of 
each other. This is a singular
situation which would presumably break supersymmetry as in the case of 
D4 branes between perpendicular NS5 and D6 branes. We have found further
evidence for the extended s-rule. This additional evidence
comes from considering the M theory curve. In particular we count 
the number of complex spheres which decouple from the curve and are free to 
slide along the D6 branes or along the D6$'$ branes.
    
Worldsheet instanton effect still play an important role in the dynamics
of D4 branes suspended between D6 and D6$'$ branes. The contribution from the
D4 branes suspended between D6$'$ branes is given by taking the
s-rule between the leftmost NS5 brane and the D6$'$ branes into account  i.e. 
\begin{equation}
2 \sum_{i=1}^{N_c} i + 2 N_c ( n - N_c - 1). 
\end{equation}
The contribution of the D4 branes suspended between D6 branes
is obtained by restricting to a single D4 brane between the rightmost
D6$'$ brane and each of the D6 branes and in addition, by considering 
the s-rule between the
rightmost NS5 brane and the D6$'$ brane. The result is
\begin{equation}
2 (m - 1) + 2 (m - 3) + \cdots + 2 (m - 2 N_c + 1) = 2 N_c (m - N_c).
\end{equation} 
By adding the contribution from the D4 branes between D6$'$ and D6 branes,
the complex dimension of the Higgs moduli space is
\begin{equation}
\label{dimrot}
N_c (N_c + 1) +2 N_c ( n - N_c) + 2 N_c (m - N_c) + N_c - 2 N_c= 
2 N_c N_f - 3 N_c^2 
\end{equation} 
 The quaternionic dimension is $N_c ( n + m ) - 3/2 N_c^2 $.

For example, consider a configuration with
two D6 branes and two D6$'$ branes from left to
right in addition to NS and NS$'$ branes.
If we start with a first D4 brane, we break it between the left NS and
the first D6 brane (contribution 0), between the two D6 brane (contribution
2), between the second D6 and the first D6$'$ brane (contribution 1),
between the two D6$'$ branes (contribution 2) and between the second D6$'$
brane and the right NS brane (contribution 0).  So the total dimension is
5, which is equal to $2 \times 1 \times 4 - 3 \times 1^2 = 5$.
If we want to insert a second D4 brane, we encounter several
restrictions. We need firstly to suspend it between the left NS brane and
the second D6 brane because of the s-rule between the left NS and the the
first D6 brane. The worldsheet instanton effect tells us that the D4 brane
has then to be broken next between the second D6 brane and the second D6$'$
brane and then between the second D6$'$ brane and the right NS brane. But we
already have one D4 brane between these two and is impossible to have a
second one. Therefore it is impossible to insert a second D4 brane.
We can have only one D4 brane for this configuration. The same discussion
goes for more complicated configurations.

The following observation will help our understanding of the field theory
calculation. The D4 branes break on the $m$ D6 branes
as in any ${\cal N} = 2$ theory with NS(12345) branes and perpendicular D6
branes, the role of the right NS branes being played by the leftmost D6$'$
brane. The worldsheet instanton effect plays the role of the s-rule,
limiting the number of D4 branes between the D6 branes and the rightmost 
to one. The D4 branes break on the $n$ D6$'$ branes as in any 
${\cal N} = 1$ theory with NS, NS$'$ branes and D6 branes parallel to the 
NS$'$ branes. The role of the NS$'$ brane is played by the first $N_c$ D6
branes because we can have only one D4 brane between them and the 
leftmost D6$'$ brane. This is the case obtained if we decide to break the D4
branes first on the D6 branes. If we decide to break the D4 branes firstly
on the D6$'$ branes, the they do it on D6$'$ as in ${\cal N} = 2$ and on D6 as
in ${\cal N} = 1$ theory.

We want to turn to field theory calculations. 
Now we need to take care of the order in which we 
introduce flavor given by the D6 and the D6$'$ branes. 
There is a superpotential
coupling the adjoint field to both types of flavor given by the
D6 branes and the D6$'$ branes. 
To calculate the dimension of the moduli space we need to make the
following observation. If we introduce only D6 branes or only D6$'$ branes,
the theory would be
 ${\cal N} = 2$ supersymmetric. Let us consider that we have 
$m$ flavors given by $m$ D6 branes that break completely the gauge group. The
dimension of the moduli space is
$2 m N_c - 2 N_c^2$. Now if we introduce $n$ D6$'$ branes,  the supersymmetry 
would be 
broken to ${\cal N} = 1$. The contribution to the dimension of the moduli
space is $2 n N_c - N_c^2$ where we used the fact that for ${\cal N} = 2$
matter multiplets there are two complex scalars eaten in Higgs mechanism
whereas for ${\cal N} = 1$ matter multiplets there is only one.
By adding the two contributions to the Higgs moduli space we obtain
$2 (m + n) N_c - N_c^2$ or $2 N_f N_c - 3 N_c^2$.
Therefore, the results from the field theory and the brane configuration 
agree.
\section{Resolution of the Singularity and the Higgs Branch}
\setcounter{equation}{0}
As mentioned in the last section, 
the transition to the Higgs branch occurs when the fivebrane intersects with 
the D6-branes. This is possible when $e_i = d_i= 0$ and the Seiberg-Witten 
curve passes through the singular point $x=y=v=w=0$.
Under these conditions the term $ B_{N_c}(v, u_k)$ factorizes into
\bea
\label{}
 B_{N_c}(v, u_k) = v^r( v^{N_c -r} + \cdots + u_{N_c -r}),
\eea
where $r>0$ and $u_{N_c -r} \not = 0$.
Notice that the threefold will be of the form
$$
f(x,y,v,w) := xy - v^{N_f}w^{N_f'} =0.
$$


As a warm-up exercise, 
in this section, we show how to resolve a quadratic threefold
singularity  
\bea
xy = v w.
\eea
As explained below, the resolution is not unique. The relationship between
the different resolutions is discussed.

{$\bullet$  Type A Blow-up}

First, we blow up the $x-y-v-w$ space at $x=v=0$ by replacing
the $x-y-v-w$ space by a union of two spaces - coordinatized by
$(x, y, \tilde{v} ,w)$ and $(\tilde{x}, y, v, w)$ -which are mapped
to the $x-y-v-w$ space by $(x,y,v,w) = 
(x, y, x\tilde{v} ,w) = (\tilde{x} v, y,v,w)$. The $x-y-\tilde{v}-w$
and the $\tilde{x}-y-v-w$ spaces are glued together by the relation
$\tilde{x}\tilde{v} =1$ and $v=x\tilde{v}$. The equation $xy = vw$ becomes 
$x(y- \tilde{v} w)$
in the $x-y-\tilde{v}-w$ space and $v(\tilde{x}y - w)$ in the 
$\tilde{x}-y-v-w$ space. If we ignore the piece described by $x=0$ and $v=0$,
which is mapped to the $y-w$ plane $x=v=0$, we obtain a union of
two smooth threefolds - $U_1 = \{y=\tilde{v} w \}$ in the 
$x-y-\tilde{v}-w$ space and $U_2 = \{\tilde{x}y = w \}$ in the
$\tilde{x}-y-v-w$ space. The threefolds $U_1$ and $U_2$ are coordinatized by
$(x, \tilde{v}, w)$ and $(\tilde{x}, y, v)$ respectively and
glued together by the condition
$\tilde{v}\tilde{x} = 1$, \, $v=x\tilde{v}$
 and $y =\tilde{v}w$. Thus we obtain
a smooth threefold. This threefold is mapped onto the original
singular $A_{0,0}$ threefold $xy =vw$: \,\, $(x,y,v,w) = (x, \tilde{v}w, 
x\tilde{v} , w)$ on $U_1$ and $(x,y,v,w) = (\tilde{x} v, y, v, \tilde{x}y)$
on $U_2$. The inverse image of the singular point $x=y=v=w=0$ is
described by $x=w=0$ in $U_1$ and by $y=v=0$ in $U_2$. 
It is coordinatized  by $\tilde{v}$ and $\tilde{x}$ which are related by
$\tilde{v}\tilde{x}=1$ and thus, is a projective line ${\bf P}^1$.

{$\bullet$ Type B Blow-up}

Second, we blow up the $x-y-v-w$ space at $x=w=0$ by replacing
the $x-y-v-w$ space by a union of two spaces - coordinatized by
$(x, y, v,  \tilde{w})$ and $(\tilde{x}, y, v, w)$ -which are mapped
to the $x-y-v-w$ space by $(x,y,v,w) = 
(x, y, v, x\tilde{w}) = ( \tilde{x} w, y, v,w)$. If we go through the same 
process, then we will obtain a union of two smooth threefolds - 
$V_1 = \{y= v\tilde{w} \}$ 
in the 
$x-y-v-\tilde{w}$ space and $V_2 = \{\tilde{x}y = v \}$ in the
$\tilde{x}-y-v-w$ space. The threefolds $V_1$ and $V_2$ are coordinatized by
$(x, v, \tilde{w})$ and $(\tilde{x},y, w)$ respectively and
glued together by $\tilde{w}\tilde{x} = 1$, \, $w=x\tilde{w}$
and $y =v\tilde{w}$. Thus we obtain
a smooth threefold. This threefold is mapped onto the original
singular $A_{0,0}$ threefold $xy =vw$: \,\, $(x,y,v,w) = (x,v\tilde{w},
v, x\tilde{w})$ on $V_1$ and $(x,y,v,w) = (\tilde{x}w, y, \tilde{x}y,w)$
on $V_2$. The inverse image of the singular point $x=y=v=w=0$ is
described by $x=v=0$ in $V_1$ and by $y=w=0$ in $V_2$. 
It is coordinatized  by $\tilde{w}$ and $\tilde{x}$ which are related by
$\tilde{w}\tilde{x}=1$, and thus is a projective line ${\bf P}^1$.

In this special case, the resolved threefold we obtained in Type A blow-up
and Type B blow-up are isomorphic. This is not true in general. 
Note that there is no isomorphism between the type A and type B blowup
which will commute with the maps to the original singular variety.
Two resolved threefold are related by a birational transformation, called
a flop. The following diagram compares Type A and Type B blow-ups.
\newline
\\
\noindent
\begin{figure}
\begin{picture}(300, 130)(-40,0)
\setlength{\unitlength}{0.5mm}
\put(70, 80){$U_1 \cup U_2$}
\put(180, 80){$ V_1 \cup V_2$}
\put(130, 20){$X$}
\put(85, 70){\vector(1,-1){35}}
\put(190, 70){\vector(-1,-1){35}}
\put(130, 85){$\mbox{flop}$}
\thicklines
\dashline{12.000}(110,82)(130,82)(150,82)(170,82)
\end{picture}
\caption{Comparison of Type A and Type B Blow-ups}
\end{figure}
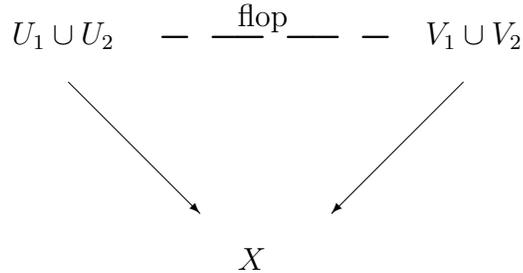
\\
In this example, the flop is just a result of changing the role of $v$
and $w$, which corresponds to the exchange of $D6$ and D6$'$ branes.

\section{Resolution of $A_{-1,n-1}$ singularity and Higgs Branch}
Before discussing the resolution of the singularity, it is helpful to
make an observation regarding the mass of the adjoint field.  
Usually the mass of the adjoint field is connected to the angle between
the NS5 branes when the D6 branes are not rotated. 
However, the mass of the adjoint field is not a function of the angle 
between NS5 branes alone:
it also depends on the angle between D6 branes and NS5 branes.
The superpotential is also a function of the angles between D6 and NS5
branes. The mass of the adjoint can be defined by measuring the strength
of the quartic superpotential of the quarks obtained after integrating the 
massive adjoint out. In this way we see that if 
we have both superpotential terms and
terms proportional to $\Phi^2$, then the mass of the adjoint
is a function of the relative angle between the 
D6 branes and the NS5 branes.\footnote{We would like to thank Amihay
Hanany and
David Kutasov for useful discussions on this matter. See \cite{gk} for
a complete discussion.} In this section we consider the case 
involving NS5 branes and D6$'$ branes which are parallel. This means 
that the mass of the adjoint is zero and there is no superpotential 
in the ${\cal N} = 1$ theory.   

We now turn to the problem of resolving
the singularity of a threefold ${\cal X} \subset {\bf C}^4$ given by
\bea
\label{ref}
{\cal X}: xy =w^n.
\eea
which is a singularity of type $A_{-1, n-1}$.

We blow up ${\bf C}^4$ -coordinatized by $(x,y,v,w)$
along the plane  $y=w=0$ by replacing
the $x-y-v-w$ space (i.e. ${\bf C}^4$)
  by a union of two spaces - coordinatized by
$(x, y, v,  \tilde{w})$ and $(x, \tilde{y}, v, w)$ -which are mapped
to the $x-y-v-w$ space 
by $(x,y,v,w) = (x, y, v, y\tilde{w}) = (x, \tilde{y} w, v,w)$. 
This corresponds to surgery along the plane $y=w=0$ which replaces
the plane by the product of a  plane and a projective line ${\bf P}^1$. 
 The $x-y-v-\tilde{w}$
and the $x-\tilde{y}-v-w$ spaces are glued together by 
the condition $\tilde{y}\tilde{w} =1$ and
$w=y\tilde{w}$. Consider first the simplest type of singularity
i.e. $A_{-1, 1}$.
The equation $xy = w^2$ becomes $y(x- \tilde{w}^2 y)$
in the $x-y-v-\tilde{w}$ space and $w(x\tilde{y} - w)$ in the 
$x-\tilde{y}-v-w$ space. If we ignore the piece described by $y=0$ and $w=0$
which is mapped to the $x-v$ plane $y=w=0$, we obtain a union of
two smooth threefolds - $W_1 = \{x=\tilde{w}^2y \}$ in the 
$x-y-\tilde{v}-w$ space and $W_2 = \{x\tilde{y} = w \}$ in the
$x-\tilde{y}-v-w$ space. The threefolds $W_1$ and $W_2$ are coordinatized by
$(y, v, \tilde{w})$ and $(x, \tilde{y}, v)$ respectively and
glued together by $\tilde{w}\tilde{y} = 1$, \, $x=y\tilde{w}^2$
 and $w =x\tilde{y}$. Thus we obtain
a smooth threefold. This threefold is mapped onto the original
singular  threefold $xy =w^2$: \,\, $(x,y,v,w) = (\tilde{w}^2 y, 
y,v, y\tilde{w})$ on $W_1$ and 
$(x,y,v,w) = (x, x \tilde{y}^2, v, x\tilde{y})$
on $W_2$. The inverse image of the singular line $x=y=w=0$ is
described by $y=0$ in $W_1$ and by $x=0$ in $W_2$. 
It is coordinatized  by $\tilde{w}, v$ and $\tilde{y}, v$ which are related by
$\tilde{w}\tilde{y}=1$, and thus is a product ${\bf P}^1 \times
{\bf A}^1$ of a projective line and an affine line.

If we started with a higher $A_{-1,n-1}$ singularity, the equation 
$xy = w^n$ becomes $x= y^{n-1}\tilde{w}^n$ in the $x-y-v-\tilde{w}$ space
and $x\tilde{y} = w^{n-1}$  in the $x-\tilde{y}-v-w$ space (again ignoring the
trivial piece $y=0$ and $w=0$). 
This is smooth in the $x-y-v-\tilde{w}$ space
but has an $A_{-1, n-2}$ singularity at $x=\tilde{y}=v=w=0$
in the $x-\tilde{y}-v-w$ space. 
Thus the threefold is not yet resolved but it has become
less singular. We can further decrease $n-1$ by one by blowing up
the $\tilde{y}-w$ plane at $\tilde{y} = w =0$.
Iterating this process, we can finally resolve the singular $A_{-1,n-1}$
singularity. It is straightforward to see that the resolved space is covered
by $n$ three-space $V_1, V_2, V_3, \ldots , V_n$ with
coordinates $(y_1, w_1, v) = (y, \tilde{w}, v)$, $(y_2 = \tilde{y}, w_2, v),
(y_3, w_3, v), \ldots , (y_n, w_n = x, v)$ which are mapped to
the singular $A_{-1, n-1}$ threefold by
\bea
\label{res-map}
V_j \ni (y_j, w_j, v) \mapsto\left\{ \begin{array} {l}
	x = y_j^{n-j}w_j^{n+1 -j}\\
	y= y_j^jw_j^{j-1}\\
	v=v\\
	w= y_jw_j 
\end{array} \right.
\eea
The three-spaces $V_j$ are glued together by $w_jy_{j+1} =1$ and 
$y_jw_j = y_{j+1}w_{j+1}$. The map onto the singular $A_{-1,n-1}$ threefold
is isomorphic except at the inverse image of the singular line $x=y=w=0$.
The inverse image consists of $n-1$ ${\bf P}^1 \times {\bf A}^1$'s
$D_1, D_2, \ldots , D_{n-1}$ where $D_j$ is the locus of $y_j =0$
in $V_j$ and $w_{j+1} = 0$ in $V_{j+1}$, and is coordinatized by $w_j$ and
$y_{j+1} $ that are related by $w_jy_{j+1} =1$. $D_j$ and $D_k$ do not 
intersect unless $k= j\pm 1$, and $D_{j-1}$ and $D_j$ intersect transversely
at $y_j =w_j =0$. Inside each of the $D_j$, the projective line ${\bf P}^1$ 
defined by $v=0$ is the inverse image of the origin $x=y=v=w=0$ 
which will be
denoted by $C_j$. We denote the resolved threefold by $\widetilde{\cal{X}}$ and
the map onto the singular threefold, which is described in (\ref{res-map}), by
$\sigma$.

Thus, we have
\bea
\sigma:{\widetilde{\cal X}}& = &V_1 \bigsqcup_{w_1y_2=1} V_2 
\bigsqcup_{w_2y_3 =1} \cdots \bigsqcup_{w_{n-1}y_n=1} V_n 
\longrightarrow {\cal X}\\
\sigma^{-1}(0) &=& C_1 \cup C_2 \cup \cdots \cup C_{n-1}
\eea
and the map $\sigma$ is onto and isomorphic outside the line defined by 
$y=w=0$ on ${\cal X}$.
Consider a Seiberg-Witten curve on the singular threefold ${\cal X}$
given by 
\bea
x +y = v^r\\
w =0.
\eea
We would like to study the total transform of the curve on the resolved
threefold, which is the inverse image (in an algebraic sense) of the
curve under the map $\sigma$. On the $j-$th patch $V_j$, the equation of the
 curve will be
\bea
\label{ffirst}
 y_j^{n-j}w_j^{n+1 -j} + y_j^jw_j^{j-1} =v^r\\
\label{ssecond}
y_jw_j=0.
\eea
Thus by plugging (\ref{ssecond}) into (\ref{ffirst})
we obtain
\bea
y& =& v^r,\quad yw_1 =0\quad  \mbox{for}\quad  j=1\nonumber\\
v^r& =&0,\quad y_jw_j =0\quad  \mbox{for}\quad   2\leq j \leq n-1\nonumber\\
x&=&v^r\quad y_nx =0\quad  \mbox{for}\quad  j=n.
\eea
Thus on the $j-$th patch $V_j$ for $2\leq j \leq n-1$, the total transform of the curve
consists of $r$-multiple copies  of $C_j\cap V_j$, which is given
by $v^r=0, y_j=0$  and $r$-multiple copies  of
$C_{j-1}\cap V_j$, which is given by $v^r=0, w_j=0$. 
On $V_1$, the total transform
consists of $r$-multiples of $C_1 \cap V_1$ and a curve defined
by $y-v^r = w_1 =0$. On $V_n$, the total transform consists of
$r$-multiple copies of $C_{n-1} \cap V_n$ and a curve defined
by $x-v^r =y_n=0$. Thus the total transform
consist of $r$-multiple copies of $C_1, C_2, \ldots , C_{n-1}$ and two irreducible
curves $C_L$ and $C_R$ which  meet $C_1$ and $C_{n-1}$ tangentially but
in transversal direction.
Since each $C_j$ is isomorphic to ${\bf P}^1$ and the motion between
D6$'$ branes are parameterized by these ${\bf P}^1$'s, 
the quaternionic dimension of the r-th Higgs branch will be
 $r(n-1)$. If we now add the contribution from the D4 branes ending on
both NS5
branes (which is not counted in the above derivation) the total dimension
is $r n$. The Higgs branch is depicted in Figure 2.

In field theory we obtain the result:
\begin{equation}
2 r n + n^2 - n^2 = 2 r n
\end{equation}
i.e. a $r n$ quaternionic dimension. 

We thus have perfect agreement with the field theory results.
\begin{figure}
\setlength{\unitlength}{0.00065in}
\begingroup\makeatletter\ifx\SetFigFont\undefined%
\gdef\SetFigFont#1#2#3#4#5{%
  \reset@font\fontsize{#1}{#2pt}%
  \fontfamily{#3}\fontseries{#4}\fontshape{#5}%
  \selectfont}%
\fi\endgroup%
\begin{picture}(9016,1622)(0,-150)

\thicklines
\path(1808,8)	(1817.454,49.749)
	(1826.835,90.626)
	(1836.146,130.638)
	(1845.390,169.792)
	(1854.571,208.094)
	(1863.692,245.552)
	(1881.770,317.961)
	(1899.649,387.074)
	(1917.358,452.946)
	(1934.924,515.632)
	(1952.375,575.188)
	(1969.738,631.666)
	(1987.041,685.124)
	(2004.310,735.615)
	(2021.574,783.195)
	(2038.860,827.919)
	(2056.196,869.841)
	(2073.608,909.016)
	(2091.125,945.500)
	(2126.582,1010.612)
	(2162.785,1065.617)
	(2199.956,1110.954)
	(2238.312,1147.062)
	(2278.076,1174.382)
	(2319.465,1193.352)
	(2362.700,1204.411)
	(2408.000,1208.000)

\path(2408,1208)	(2453.300,1204.411)
	(2496.535,1193.352)
	(2537.924,1174.382)
	(2577.688,1147.062)
	(2616.044,1110.954)
	(2653.215,1065.617)
	(2689.418,1010.612)
	(2724.875,945.500)
	(2742.392,909.016)
	(2759.804,869.841)
	(2777.140,827.919)
	(2794.426,783.195)
	(2811.690,735.615)
	(2828.959,685.124)
	(2846.262,631.666)
	(2863.625,575.188)
	(2881.076,515.632)
	(2898.642,452.946)
	(2916.351,387.074)
	(2934.230,317.961)
	(2952.308,245.552)
	(2961.429,208.094)
	(2970.610,169.792)
	(2979.854,130.638)
	(2989.165,90.626)
	(2998.546,49.749)
	(3008.000,8.000)

\path(1808,608)	(1808.000,608.000)

\path(1808,608)	(1808.000,608.000)

\path(2508,8)	(2517.454,49.749)
	(2526.835,90.626)
	(2536.146,130.638)
	(2545.390,169.792)
	(2554.571,208.094)
	(2563.692,245.552)
	(2581.770,317.961)
	(2599.649,387.074)
	(2617.358,452.946)
	(2634.924,515.632)
	(2652.375,575.188)
	(2669.738,631.666)
	(2687.041,685.124)
	(2704.310,735.615)
	(2721.574,783.195)
	(2738.860,827.919)
	(2756.196,869.841)
	(2773.608,909.016)
	(2791.125,945.500)
	(2826.582,1010.612)
	(2862.785,1065.617)
	(2899.956,1110.954)
	(2938.312,1147.062)
	(2978.076,1174.382)
	(3019.465,1193.352)
	(3062.700,1204.411)
	(3108.000,1208.000)

\path(3108,1208)	(3153.300,1204.411)
	(3196.535,1193.352)
	(3237.924,1174.382)
	(3277.688,1147.062)
	(3316.044,1110.954)
	(3353.215,1065.617)
	(3389.418,1010.612)
	(3424.875,945.500)
	(3442.392,909.016)
	(3459.804,869.841)
	(3477.140,827.919)
	(3494.426,783.195)
	(3511.690,735.615)
	(3528.959,685.124)
	(3546.262,631.666)
	(3563.625,575.188)
	(3581.076,515.632)
	(3598.642,452.946)
	(3616.351,387.074)
	(3634.230,317.961)
	(3652.308,245.552)
	(3661.429,208.094)
	(3670.610,169.792)
	(3679.854,130.638)
	(3689.165,90.626)
	(3698.546,49.749)
	(3708.000,8.000)

\path(6008,8)	(6017.454,49.749)
	(6026.835,90.626)
	(6036.146,130.638)
	(6045.390,169.792)
	(6054.571,208.094)
	(6063.692,245.552)
	(6081.770,317.961)
	(6099.649,387.074)
	(6117.358,452.946)
	(6134.924,515.632)
	(6152.375,575.188)
	(6169.738,631.666)
	(6187.041,685.124)
	(6204.310,735.615)
	(6221.574,783.195)
	(6238.860,827.919)
	(6256.196,869.841)
	(6273.608,909.016)
	(6291.125,945.500)
	(6326.582,1010.612)
	(6362.785,1065.617)
	(6399.956,1110.954)
	(6438.312,1147.062)
	(6478.076,1174.382)
	(6519.465,1193.352)
	(6562.700,1204.411)
	(6608.000,1208.000)

\path(6608,1208)	(6653.300,1204.411)
	(6696.535,1193.352)
	(6737.924,1174.382)
	(6777.688,1147.062)
	(6816.044,1110.954)
	(6853.215,1065.617)
	(6889.418,1010.612)
	(6924.875,945.500)
	(6942.392,909.016)
	(6959.804,869.841)
	(6977.140,827.919)
	(6994.426,783.195)
	(7011.690,735.615)
	(7028.959,685.124)
	(7046.262,631.666)
	(7063.625,575.188)
	(7081.076,515.632)
	(7098.642,452.946)
	(7116.351,387.074)
	(7134.230,317.961)
	(7152.308,245.552)
	(7161.429,208.094)
	(7170.610,169.792)
	(7179.854,130.638)
	(7189.165,90.626)
	(7198.546,49.749)
	(7208.000,8.000)

\path(8408,1208)	(8360.442,1196.970)
	(8313.884,1186.063)
	(8268.318,1175.277)
	(8223.736,1164.607)
	(8180.131,1154.050)
	(8137.494,1143.603)
	(8095.819,1133.263)
	(8055.097,1123.025)
	(8015.321,1112.887)
	(7976.483,1102.845)
	(7901.590,1083.034)
	(7830.356,1063.567)
	(7762.720,1044.416)
	(7698.619,1025.552)
	(7637.992,1006.949)
	(7580.777,988.579)
	(7526.912,970.414)
	(7476.336,952.428)
	(7428.986,934.592)
	(7384.802,916.879)
	(7343.720,899.262)
	(7270.619,864.206)
	(7209.188,829.203)
	(7158.933,794.034)
	(7119.360,758.478)
	(7070.283,685.330)
	(7059.789,647.297)
	(7058.000,608.000)

\path(7058,608)	(7066.936,554.412)
	(7087.598,503.994)
	(7120.701,456.527)
	(7166.957,411.791)
	(7227.082,369.566)
	(7301.789,329.633)
	(7344.834,310.457)
	(7391.793,291.772)
	(7442.754,273.549)
	(7497.807,255.762)
	(7557.042,238.384)
	(7620.547,221.386)
	(7688.411,204.741)
	(7760.725,188.422)
	(7798.578,180.376)
	(7837.577,172.401)
	(7877.732,164.494)
	(7919.056,156.651)
	(7961.559,148.869)
	(8005.252,141.144)
	(8050.147,133.473)
	(8096.254,125.853)
	(8143.585,118.280)
	(8192.151,110.751)
	(8241.964,103.261)
	(8293.033,95.809)
	(8345.372,88.390)
	(8398.989,81.001)
	(8453.898,73.638)
	(8510.108,66.298)
	(8567.632,58.978)
	(8626.479,51.675)
	(8686.663,44.384)
	(8748.192,37.102)
	(8811.080,29.826)
	(8875.336,22.553)
	(8940.972,15.279)
	(9008.000,8.000)

\path(608,1208)	(655.216,1196.964)
	(701.439,1186.052)
	(746.678,1175.261)
	(790.939,1164.586)
	(834.230,1154.024)
	(876.559,1143.573)
	(917.934,1133.228)
	(958.362,1122.986)
	(997.852,1112.843)
	(1036.409,1102.797)
	(1110.762,1082.980)
	(1181.480,1063.506)
	(1248.627,1044.348)
	(1312.263,1025.479)
	(1372.449,1006.871)
	(1429.248,988.497)
	(1482.719,970.330)
	(1532.925,952.341)
	(1579.928,934.503)
	(1623.787,916.790)
	(1664.565,899.173)
	(1737.122,864.118)
	(1798.090,829.119)
	(1847.959,793.956)
	(1887.221,758.411)
	(1935.882,685.290)
	(1946.264,647.276)
	(1948.000,608.000)

\path(1948,608)	(1939.057,554.366)
	(1918.455,503.908)
	(1885.483,456.408)
	(1839.433,411.644)
	(1779.591,369.398)
	(1705.250,329.450)
	(1662.419,310.269)
	(1615.697,291.580)
	(1564.994,273.355)
	(1510.223,255.568)
	(1451.293,238.190)
	(1388.116,221.194)
	(1320.604,204.553)
	(1248.668,188.239)
	(1211.012,180.196)
	(1172.218,172.225)
	(1132.273,164.322)
	(1091.166,156.483)
	(1048.887,148.706)
	(1005.423,140.986)
	(960.765,133.321)
	(914.901,125.707)
	(867.820,118.140)
	(819.511,110.617)
	(769.962,103.135)
	(719.163,95.690)
	(667.102,88.279)
	(613.769,80.898)
	(559.152,73.544)
	(503.240,66.213)
	(446.022,58.902)
	(387.487,51.608)
	(327.625,44.328)
	(266.423,37.056)
	(203.870,29.791)
	(139.956,22.529)
	(74.670,15.267)
	(8.000,8.000)

\path(3408,8)	(3414.403,49.016)
	(3420.792,89.182)
	(3427.169,128.507)
	(3433.537,166.997)
	(3446.259,241.499)
	(3458.981,312.744)
	(3471.726,380.786)
	(3484.516,445.679)
	(3497.374,507.481)
	(3510.324,566.244)
	(3523.388,622.024)
	(3536.588,674.876)
	(3549.949,724.855)
	(3563.493,772.016)
	(3577.242,816.413)
	(3591.220,858.102)
	(3605.449,897.138)
	(3619.953,933.575)
	(3649.874,998.874)
	(3681.168,1054.437)
	(3714.018,1100.706)
	(3748.605,1138.119)
	(3785.114,1167.115)
	(3823.727,1188.134)
	(3864.628,1201.616)
	(3908.000,1208.000)

\path(3908,1208)	(3954.844,1207.398)
	(4000.079,1198.926)
	(4043.962,1182.147)
	(4086.748,1156.619)
	(4128.695,1121.905)
	(4170.058,1077.563)
	(4211.094,1023.156)
	(4252.060,958.242)
	(4272.596,921.709)
	(4293.211,882.384)
	(4313.936,840.213)
	(4334.804,795.141)
	(4355.847,747.113)
	(4377.096,696.075)
	(4398.583,641.970)
	(4420.342,584.744)
	(4442.403,524.343)
	(4464.799,460.711)
	(4487.562,393.794)
	(4510.724,323.536)
	(4534.316,249.882)
	(4546.284,211.765)
	(4558.372,172.779)
	(4570.583,132.916)
	(4582.922,92.169)
	(4595.393,50.533)
	(4608.000,8.000)

\put(4808,608){$\bullet \quad \bullet \quad \bullet$}
\put(2078,1448){$C_1$}
\put(2918,1463){$C_2$}
\put(3758,1448){$C_3$}
\put(6443,1433){$C_{n-1}$}
\put(548,1433){$C_L$}
\put(8303,1418){$C_R$}
\end{picture}
\caption{The Higgs Branch}
\end{figure}
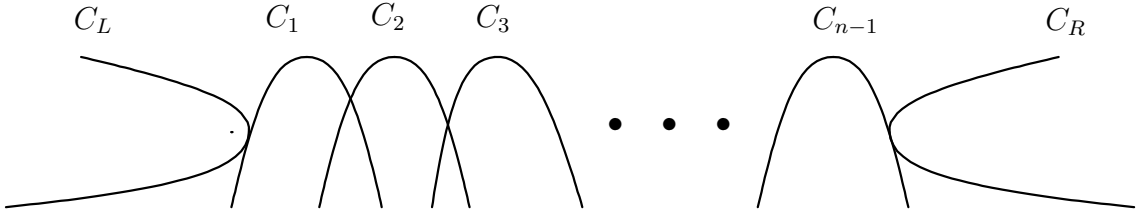

\section{Resolution of $A_{m-1, n-1}$ singularity}
A complex 3-dimensional hypersurface singularity defined by
\bea
\label{amn}
f(x,y,v,w) := xy - v^m w^n =0
\eea 
in ${\bf C}^4$ will be called a type $A_{m-1, n-1}$ singularity.
For $m>1$ and $n>1$, the $A_{m-1, n-1}$ singularity
 will be singular along the union of two complex lines
 $x=y=v=0$ and $x=y=w=0.$ This is easily seen by noting that
there is no well-defined normal vector to the threefold $xy = v^m w^n$
along these lines while outside of these lines either 
$\partial f/ \partial x$ or $\partial f/ \partial y$ will provide
a normal vector to the threefold.  Thus the singularity is not isolated 
unless $m=n=1$. 

We can resolve the $A_{m-1, n-1}$ singularity 
by successive (small) blow-ups. 
However there are many possible desingularizations
which are smooth threefolds mapping surjectively onto the threefold (\ref{amn})
and isomorphically  over the smooth points of (\ref{amn}).
Hence they are birational and related to each other by flop transitions. 

We now consider the problem of resolving the $A_{m-1,n-1}$ 
singularity for higher $m$ and $n$.
We first use Type A blow-up i.e. blowing up the $x-y-v-w$ space at $x=v=0$.
Then the equation 
$xy =v^m w^n$ becomes $y=x^{m-1}\tilde{v}^m w^n$ in 
the $x-y-\tilde{v}-w$ space
and $\tilde{x}y = v^{m-1}w^n$  in the $\tilde{x}-y-v-w$ space (once again,
ignoring the
trivial piece $x=0$ and $v=0$). It is smooth in the $x-y-\tilde{v}-w$ space
but has an $A_{m-2, n-1}$ singularity 
at $\tilde{x}=y=v=w=0$ in the $\tilde{x}-y-v-w$ space. 
Thus the threefold is not yet resolved, but it has become
less singular. We can further decrease $m-1$ by one by blowing up
the $\tilde{x}-v$ plane at $\tilde{x} = v=0$.
Iterating this process, we arrive  the singularity  of type $A_{-1,m-1}$
in $x_m-y-v-w$ space . Now by applying Type B blow-up successively
i.e. blowing up the 
$x_m-y-v-w$ space at $x_m=w=0$, we can finally resolve the singularity. 
It is straightforward to see that the resolved space is covered
by $n +m$ three-spaces $U_1, U_2, U_3, \ldots , U_m, V_1, V_2, \ldots,
V_n$ with
coordinates  $(x_1, v_1, w) = (x, \frac{v}{x}, w)$, $(x_2,  
v_2, w)=(\frac{x_1}{v}, \frac{v}{x_2}, w) ,\ldots ,
(x_i, v_i, w) = (\frac{x_{i-1}}{v}, \frac{v}{x_i}, w) , 
\ldots , (x_m, v_m, w)=(\frac{x_{m-1}}{v}, \frac{v}{x_{m}},w),$
$(x_{m+1}, w_1, v) = (\frac{x_m}{v}, \frac{w}{x_{m+1}}, v)$, 
$(x_{m+2}, w_2,v)=  ( \frac{x_{m+1}}{w}, \frac{w}{x_{m+2}}, v), \ldots
(x_{m+j}, w_j, v) =(\frac{x_{m+j-1}}{w}, \frac{w}{x_{m+j}}, v),
\ldots , (x_{m+n}, w_n = y, v)$ which are mapped to
the singular $A_{m-1, n-1}$ threefold by
\bea
\label{gen-res}
U_i \ni (x_i, v_i, w) \mapsto\left\{ \begin{array} {l}
	x = x_i^iv_i^{i-1}\\
	y=x_i^{m-i}v_i^{m+1 -i} w^n\\
	v=x_iv_i\\
	w= w 
\end{array} \right. \\
V_j \ni (x_{m+j}, w_j, v) \mapsto\left\{ \begin{array} {l}
	x = x_{m+j}^{j}w_j^{j-1}v^{m}\\
	y= x_{m+j}^{n-j}w_j^{n+1-j}\\
	v=v\\
	w= x_{m+j}w_j.
\end{array} \right.
\eea
The three-spaces $U_i$ are glued together by $v_ix_{i+1} =1$ and 
$x_iv_i = x_{i+1}v_{i+1}$ and the three-spaces $V_j$ are glued together 
by $w_jx_{m+j+1} =1$ and 
$x_{m+j}w_j = x_{m+j+1}w_{j+1}$. Finally, the three-spaces $U_m$ and $V_1$ are
glued together by $x_{m+1} v_m= 1, w=w_1 x_{m+1}, v=v_mx_m$.

We denote the resolved threefold by $\tilde{\cal{X}}$ and
the map onto the singular threefold by $\sigma$ which is described by
(\ref{gen-res}).

$\sigma$  maps  the resolved threefold onto
the singular $A_{m-1, n-1}$ threefold isomorphically 
outside  the singular locus, which is a union of
two lines defined by $x=y=v=0$ and $x=y=w=0$ on ${\cal X}$. We will now
consider
the exceptional loci of $\sigma$ i.e. the set of points where $\sigma$
is not one-to-one. There are three types of exceptional loci
corresponding to the movement of D4 branes between two D6$'$ branes,
between two D6 branes and finally between D6$'$ and D6 branes. 

The inverse image of the singular line $x=y=v=0$
consists of $(m-1)$ ${\bf P}^1 \times {\bf A}^1$'s
$B_1, B_2, \ldots , B_{m-1}$ where $B_i$ is the locus of $x_i =0$
in $U_i$ and $v_{i+1} = 0$ in $U_{i+1}$, and is coordinatized by $v_i$ and
$x_{i+1} $ that are related by $v_ix_{i+1} =1$. 
$B_i$ and $B_j$ do not 
intersect unless $j = i\pm 1$, and $B_{i-1}$ and $B_i$ intersect transversely
at $x_i =v_i =0$. Inside of each $B_i$, the projective line ${\bf P}^1$ 
defined by $w=0$ is the inverse image of the origin $x=y=v=w=0$ 
which will be
denoted by $A_i$. These ${\bf P}^1$'s correspond to the 
D4 branes suspended between two D6$'$ branes in the Higgs branch.

The inverse image of the singular line $x=y=w=0$
consists of $(n-1)$ 
${\bf P}^1 \times {\bf A}^1$'s
$D_1, D_2, \ldots , D_{n-1}$ where $D_j$ is the locus of $x_{m+j} =0$
in $V_j$ and $w_{j+1} = 0$ in $V_{j+1}$, and is coordinatized by $w_j$ and
$x_{m+j+1} $ that are related by $w_jx_{m+j+1} =1$.
$D_j$ and $D_k$ do not 
intersect unless $k= j\pm 1$, and $D_{j-1}$ and $D_j$ intersect transversely
at $y_j =w_j =0$. Inside of each $D_j$, the projective line ${\bf P}^1$ 
defined by $v=0$ is the inverse image of the origin $x=y=v=w=0$ 
which will be
denoted by $C_j$. These ${\bf P}^1$'s correspond to the 
D4 branes between two D6 branes in the Higgs branch.

Finally there is a special exceptional divisor ${\bf P}^1$, denoted by $E$,
which is given by $x_m=w=0$ in $U_m$ and $w_1=v=0$ in $V_1$ in the sense that
$E$ does not move while $A_i$ and $C_j$ can move in a family 
${\bf P}^1\times {\bf A}^1$ of ${\bf P}^1$'s.
$E$ corresponds to the D4 branes suspended between D6 and D6$'$ branes
in the Higgs branch.

Thus we have
\bea
\sigma:{\tilde{\cal X}}& = &U_1 \cup  U_2  \cdots \cup
U_m \cup  V_1 \cup  V_2 
\cdots \cup V_n 
\longrightarrow {\cal  X}\\
\sigma^{-1}(0) &=& A_1 \cup A_2 \cup \cdots \cup A_{m-1}\cup E \cup
 C_1 \cup C_2 \cup \cdots \cup C_{n-1}.
\eea

Consider now, a Seiberg-Witten curve on the singular threefold ${\cal X}$
in a general position in $v-w$ space
given by 
\bea
\label{sw}
x +y = (v+\mu w)^r\\
\mu v - w =0
\eea
where $\mu\neq 0, \infty$ is a constant. In the conventions described
above equation (\ref{ref}) this would correspond to a configuration 
with parallel NS5 branes, rotated at angle $\mu = tan(\theta)$
with respect to the D6$'$ branes.

We would like to study the total transform of the curve on the resolved
threefold, which is the inverse image (in an algebraic sense) of the
curve under the map $\sigma$. 
On the $i-$th patch $U_i$, the equation of the curve will be
\bea
\label{gU_i-eqn1}
x_i^iv_i^{i-1} + x_i^{m-i}v_i^{m+1 -i} w^n = ( x_iv_i + \mu w)^r\\
\mu v - w=0.
\eea
By plugging the second equation into the first equation, we have
\bea
\label{gredU_i-eqn1}
x_i^iv_i^{i-1} + \mu^n x_i^{m+n -i}v_i^{m+n+1-i} = (1+ \mu^2)^r  x_i^rv_i^r.
\eea
The equation (\ref{gredU_i-eqn1}) will factorize into
\bea
x_i^iv_i^{i-1}(1  + \mu^n x_i^{m+n -2i}v_i^{m+n-2i+2} 
- (1+\mu^2)^rx_i^{r-i}v_i^{r+1-i}) =0,\quad i=1, \ldots , r,\nonumber\\
x_i^rv_i^r(x_i^{i-r}v_i^{i-1-r} +  \mu^n x_i^{m+n -r-i}v_i^{m+n-r-i+1}
-(1+\mu^2)^r) =0,\quad i = r+1, \ldots , m.
\eea
Thus on each $i-$th patch $U_i$ for $ i=1, \ldots , r$, 
the total transform of the curve
consists of $i$-multiple copies  of $A_i\cap U_i$, which is given
by $x_i^i=0, w=0$  and $(i-1)$-multiple copies  of
$A_{i-1}\cap U_i$, which is given by $v_i^{i-1}=0, w=0$. 
On each $i-$th patch $U_i$ for $ i=r+1, \ldots , m$, 
the total transform
consists of $r$-multiples of $A_i \cap U_i$ and a curve defined
by $x_i^r= w =0$ and  $r$-multiples of $A_{i -1}\cap U_i$ and a curve defined
by $v_i^r= w =0$. 

On the other hand, the equation of the
 curve on the $j-$th patch $V_j$ will be
\bea
\label{irst}
x_{m+j}^{j}w_j^{j-1}v^{m} + x_{m+j}^{n-j}w_j^{n+1-j}
=(v+\mu  x_{m+j}w_j )^r\\
\label{econd}
\mu v -x_{m+j}w_j =0.
\eea
Thus by plugging (\ref{econd}) into (\ref{irst})
the equation can be rewritten as
\bea
\mu^{-m}x_{m+j}^{m+j}w_j^{m+j-1}
 + x_{m+j}^{n-j}w_j^{n+1-j}  =(\mu^{-1}+ \mu)^rx_{m+j}^rw_j^r.
\eea
This will factorize into
\bea
x_{m+j}^rw_j^r((\mu^{-1} +\mu)^r - \mu^{-m} x_{m+j}^{m+j-r}w_j^{m+j-1-r}
-x_{m+j}^{n-j-r}w_j^{n+1-j-r}) = 0 \nonumber\\
\quad {\mbox {for}} \quad j =1, \ldots , n-r,\nonumber\\
x_{m+j}^{n-j}w_j^{n+1-j}(\mu^{-m}x_{m+j}^{m-n+2j}w_j^{m-n+2j-2} + 1 -
(\mu^{-1}+\mu)^r x_{m+j}^{r -n+j}w_j^{r-n-1+j}) =0 \nonumber\\
\quad {\mbox {for}} \quad  j=n-r+1, \ldots , n.
\nonumber\\
\eea

This is basically the same set of the equations as in (\ref{gredU_i-eqn1}).
Thus on each $j-$th patch $V_j$ for $ j=1, \ldots , n-r$, 
the exceptional component of the curve
consists of $r$-multiple copies  of $C_j\cap V_j$ defined
by $x_{m+j}^r=0, v=0$  and $r$-multiple copies  of
$C_{j-1}\cap V_j$ defined by $w_j^{r}=0, v=0$. 
On each $j-$th patch $V_j$ for $ j=n-r+1, \ldots , n$, 
the exceptional components
consist of $(n-j)$-multiples of $C_j \cap V_j$ defined
by $x_{m+j}^{n+1-j}= v =0$ and  $(n+1-j)$-multiples of $C_{j-1}\cap V_j$ 
defined
by $w_j^{n+1-j}= v =0$.

By counting the multiplicities of the complex spheres above,
we have verified that the s-rule between the D6 and NS5 branes
and between the D6$'$ and NS5 branes is naturally encoded in the 
geometry. This provides a non-trivial check of this s-rule.

The quaternionic dimension of the r-th Higgs branch will be
\bea
\label{higgs-gen}
\sum_{i=1}^{r} i + r (m-r-1)+  r (n-r) + \sum_{i=1}^{r-1} i    =
r(m+n -r-1).
\eea
We need to add $r/2$ from the contribution of the D4 branes between 
the rightmost D6$'$ brane and the D6 branes to obtain:
\bea
\label{higgs-gen1}
r (m + n - r - 1/2).
\eea

A crucial observation is in order here. If we compare the results of
(\ref{dimrot}) and (\ref{higgs-gen1}), they do not agree. What is the origin
of the mismatch between the result obtained by considering the dynamics of the
brane configuration and the result obtained from a study of the geometry?
This discrepancy can be traced back to the rule (i.e. long ranged repulsion) between
the D4 branes suspended between D6 and D6$'$ branes. This long ranged repulsion
is a worldsheet instanton correction which gives an important contribution to
the dynamics of the theory. The geometry is apparently not corrected by this 
instanton effect\footnote{We thank Angel Uranga for a very important 
discussion regarding this mismatch between the brane configuration result and
the result obtained by geometrical arguments, due to the instanton effects.}.
As explained in \cite{ah,hoo1}, these instantons correspond to Euclidean 
membranes wrapping a ${\bf P^1}$ and an interval in the $x^7$ direction between 
a pair of D4 branes. These branes can be reinterpreted as fundamental strings
with a rectangular worldsheet defined by the D6, D6$'$ and D4 branes. As 
discussed in \cite{ah} and as pointed out to us by Uranga, the Euclidean string
transforms under three dimensional mirror symmetry to an ADS instanton, which is 
known to lift the Coulomb branch. In this way, we see that the instanton (
corresponding to Euclidean fundamental strings or membranes) has a
clear affect on the Higgs moduli space. This effect can not be reproduced from
a study of the equations for the geometry. The way in which M theory accounts for
these instanton corrections is an interesting open problem.

Consider now, the Seiberg-Witten curve after moving to the special position in 
$v-w$-space reached by setting $\mu =0$ in $(\ref{sw})$.
The equation is be given by
\bea
x +y = v^r\\
w =0.
\eea
We would like to study the total transform of the curve on the resolved
threefold, which is the inverse image (in an algebraic sense) of the
curve under the map $\sigma$. If we again use the convention discussed
after equation (\ref{ref}), we see that this case corresponds to the 
situation 
when the NS5 branes are parallel to the D6$'$ branes. The mass of the
adjoint field is zero.

On the $i-$th patch $U_i$, the equation of the curve will be
\bea
\label{U_i-eqn1}
x_i^iv_i^{i-1} + x_i^{m-i}v_i^{m+1 -i} w^n = x_i^rv_i^r\\
w=0.
\eea
Thus these equations will reduce to
\bea
\label{redU_i-eqn1}
x_i^iv_i^{i-1} -x_i^rv_i^r =0\\
w=0.
\eea
The equation (\ref{redU_i-eqn1}) will factorize into
\bea
x_i^iv_i^{i-1}(1- x_i^{r-i}v_i^{r+1-i}) =0\quad i=1, \ldots , r\nonumber\\
x_i^rv_i^r(x_{i}^{i-r}v_i^{i-1-r} -1) =0\quad i = r+1, \ldots , m.
\eea
Thus on each $i-$th patch $U_i$ for $ i=1, \ldots , r$, 
the total transform of the curve
consists of $i$-multiple copies  of $A_i\cap U_i$, which is given
by $x_i^i=0, w=0$  and $(i-1)$-multiple copies  of
$A_{i-1}\cap U_i$, which is given by $v_i^{i-1}=0, w=0$. 
On each $i-$th patch $U_i$ for $ i=r+1, \ldots , m$, 
the total transform
consists of $r$-multiples of $A_i \cap U_i$ and a curve defined
by $x_i^r= w =0$ and  $r$-multiples of $A_{i -1}\cap U_i$ and a curve defined
by $v_i^r= w =0$. 

On the other hand, the equation of the
 curve on the $j-$th patch $V_j$ will be
\bea
\label{first}
x_{m+j}^{j}w_j^{j-1}v^{m}+ x_{m+j}^{n-j}w_j^{n+1-j} =v^r\\
\label{second}
x_{m+j}w_j =0.
\eea
Thus by plugging (\ref{second}) into (\ref{first})
the equation can be rewritten as
\bea
x_{m+1}v^m& =& v^r,\quad x_{m+1}w_1 =0\quad  \mbox{for}\quad  j=1\nonumber\\
v^r& =&0,\quad x_{m+j}w_j =0
\quad  \mbox{for}\quad   2\leq j \leq n-1\nonumber\\
w_n&=&v^r\quad x_{m+n}w_n =0\quad  \mbox{for}\quad  j=n.
\eea
Thus on the each $j-$th patch $V_j$ for $2\leq j \leq n-1$, the exceptional components of the 
total transform 
consist of $r$-multiple copies  of $C_j\cap V_j$ given
by $v^r=0, x_{m+j}=0$  and $r$-multiple copies  of
$C_{j-1}\cap V_j$ given by $v^r=0, w_j=0$. 
On $V_1$, the total transform
consists of $r$-multiples of $C_1 \cap V_1$ given by $x_{m+1} = v^r=0$,
$r$-multiples of $E\cap V_1$ given by $v^r=w_1=0$,
and a curve defined
by $x_{m+1}v^{m-r}-1 = w_1 =0$. On $V_n$, the total transform consists of
$r$-multiple copies of $C_{n-1} \cap V_n$ given by $w_n=v^r=0$ and a curve
given by $x_{m+n}= y -v^r =0$.. 
The quaternionic dimension of the r-th Higgs branch will thus be
\bea
\label{last}
\sum_{i=1}^{r} i + r (m-r-1) +r(n-1) =r(m+n -\frac{r+3}{2}).
\eea

We again see that the result obtained from the brane configuration
does not agree with the result obtained from geometrical arguments.
This can again be traced back to contributions from worldsheet
instantons that are not accounted for in the curve equations. 
It would be very interesting to see how these instanton effects 
could be incorporated in the theory.
\vskip 1cm
{\bf Acknowledgments}
We would like to thank Edward Witten, David Morrison, Amihay Hanany and David Kutasov
for important comments on our work. We are grateful to Angel Uranga for
discussions which helped to clarify different aspects of our work 
and for his crucial suggestions.
We would like to think our referee for thoughtful comments on the previous 
version of this paper.

\end{document}